\documentclass[prl,twocolumn]{revtex4}
\input{epsf}
\newcommand{\bcen}{\begin{center}}
\newcommand{\ecen}{\end{center}}
\newcommand{\btab}{\begin{tabular}}
\newcommand{\etab}{\end{tabular}}
\newcommand{\bdes}{\begin{description}}
\newcommand{\edes}{\end{description}}

\newcommand{\beq}{\begin{equation}}
\newcommand{\eeq}{\end{equation}}
\newcommand{\bea}{\begin{eqnarray}}
\newcommand{\eea}{\end{eqnarray}}
\newcommand{\non}{\nonumber}

\newcommand{\half}{\frac{1}{2}}
\newcommand{\bary}{\begin{array}}
\newcommand{\eary}{\end{array}}
\newcommand{\beps}{\mbox{\boldmath $ \epsilon $}}
\newcommand{\bsig}{\mbox{\boldmath $ \sigma $}}

\newcommand{\bn} { \mbox{\boldmath $n$}}

\newcommand{\bu} { \mbox{\boldmath $u$}}

\newcommand{\bI} { \mbox{\boldmath $I$}}

\newcommand{\prn}[1] {(\ref{#1})}

\newcommand{\fig}[1]{fig.~\ref{#1}}
\newcommand{\Fig}[1]{Fig.~\ref{#1}}

\usepackage{epsfig}

\newcommand{\D}[1]{\mbox{d}{#1}}
\newcommand{\uo}{u_\circ}
\newcommand{\Fo}{F_\circ}

\begin{document}
% You should use BibTeX and revtex.bst for references
\bibliographystyle{apsrev}

%\preprint{}

\title{Stability of a Thin Solid Film with Interactions}
% Optional argument for running titles on pages
%\title[]{}


% \affiliation command. The \affiliation command MUST follow the

\author{Vijay Shenoy}
\email[]{vbshenoy@iitk.ac.in}
\altaffiliation{Department of Mechanical Engineering}
\author{Ashutosh Sharma}
\email[]{ashutos@iitk.ac.in}
\altaffiliation{Department of Chemical Engineering}
\affiliation{Indian Institute of Technology Kanpur, UP 208 016, India }


\date{\today}

\begin{abstract}

We investigate the question of stability of a solid thin film which
experiences external interactions such as van der Waals forces
from a contacting surface or forces from an external electric field.
Both perfectly elastic and viscoelastic material behaviours are 
considered in linear stability analysis performed here. These analyses
indicate that for sufficiently soft (shear modulus between 1 and 10
MPa) and nearly incompressible films (Poisson's ratio close to 0.5),
bifurcations are possible, i.~e., the surface of the film becomes
non-planar. The modes of bifurcation and rates of growth of
perturbations are determined as a function of material
parameters. The results of this study are of significance in
understanding the adhesive properties between a soft material (such as
rubber) and a comparatively rigid solid (such as steel), and the
behaviour of soft solid films in an electric field.
 
\end{abstract}
%\pacs{}

%\maketitle must follow title, authors, abstract and PACS
\maketitle

Instabilities and pattern formation in thin solid and liquid films are
of interest both from a scientific and technological view
point. Morphological instabilities in thin liquid films occur due to
causes such as competition between capillary forces and van der Waals
interactions \cite{Safran1993} or an external electric field
\cite{Schaffer2000} and can often lead to dewetting leading to
interesting patterns. Morphological instabilities are also common in
solid films; for example, in stressed solid films the strain energy
drives the surface roughening in competition with the surface energy
with surface mass diffusion being the dissipative
mechanism\cite{Ruckenstein1978,Freund1994}.

Analysis of interacting thin films has hitherto been restricted to
fluid films.  Here, we pose the question of stability of a thin {\em
solid} film bonded to a rigid substrate whose free surface experiences
an effective force. This force may arise from any of the various
causes such as a van der Waals interaction with another contacting
surface nearby and/or with the substrate, an external electric field,
etc. The theoretical analysis presented in this paper indicates that
for a {\em soft and nearly incompressible} solid thin film,
instabilities are possible and that the film ``buckles''. Physically,
this instability occurs because it is possible, for sufficiently large
interaction forces, to reduce the net potential energy
of the system (the elastic strain energy and the surface energy of the
film + potential energy of interaction of the surface) by a periodic
non homogeneous deformation in the film. We believe that these results
could be useful in understanding phenomena of adhesion between
materials (such as rubber and steel), behaviour of thin films in an
electric field etc.

\begin{figure}
\centerline{\input{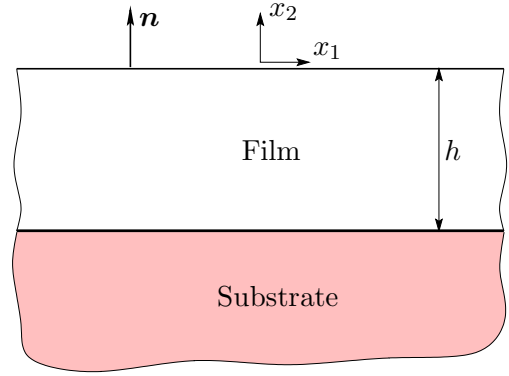}}
\caption{Film bonded to a rigid substrate. The surface of the film
experiences external forces.}
\label{scheme}
\vspace{-0.25truecm}
\end{figure}

The system considered here is shown in \fig{scheme} -- a film of
height $h$  bonded to a rigid substrate described by
coordinates $(x_1,x_2)$ such that surface of the film $S$ interacting with
external agency has $x_2 =0$ and that bonded with the rigid substrate
has $x_2 = -h$. We restrict attention to plane strain deformations of
the film for the sake of mathematical simplicity and to understand the
essential physics. The total potential energy of this system
system is
{\small 
\bea
\!\!\!\!\!\!\!\!\!\!\!\!\int_V W(\beps) \D{V} + \int_S \left(\gamma \sqrt{1+ (u_{2,1})^2} - U(\bu \cdot \bn)\right) \D{S} \label{ex_energy}
\eea
}
where $\beps$ is the strain tensor, $W(\beps)$ is the elastic strain
energy density, $\gamma$ is the surface energy, $U(\bu \cdot \bn)$ is the interaction potential
between the surface of the film and the external agency such as a
contactor or an electric field, $\bu$ is the displacement vector and $\bn$ is the
outward normal to the surface.  Linearised analysis is performed
by expanding the  interaction term $U(\bu \cdot \bn)$ in a power series
about $\bu = {\bf 0}$ and retaining all terms up to quadratic order in
$\bu$. The resulting approximate energy functional is
{\small 
\bea
& &\int_V W(\beps) \D{V} + \int_S\gamma \sqrt{1+ (u_{2,1})^2} \D{S} \non
\\
& & \;\;\;\; - \int_S \left( U_0 + \Fo \bu\cdot\bn
+ \frac{1}{2} Y (\bu \cdot \bn)^2 \right) \D{S} \label{enefunc}
\eea
}
where
{\small 
\bea
U_0 = U(0), \;\; \Fo = U'(0) \;\;\mbox{and}\;\; Y = U''(0).
\eea
}
\begin{figure}
\centerline{$\!\!\!\!\!\!\!\!\!\!\!\!\!$ \epsfxsize=7.5truecm \epsfbox[95 109 648 506]{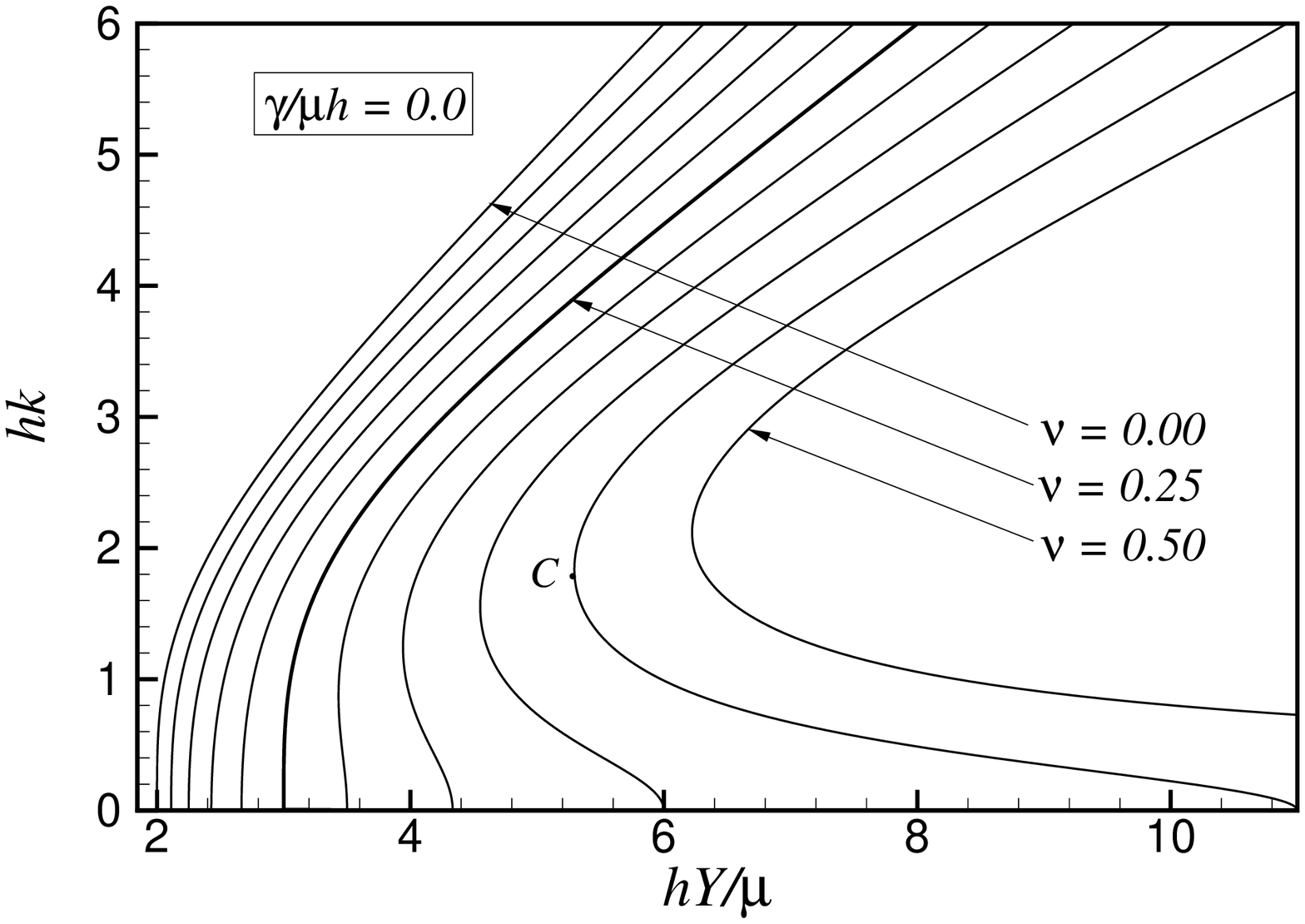}}
\centerline{(a)}
\centerline{$\!\!\!\!\!\!\!\!\!\!\!\!\!$ \epsfxsize=7.5truecm \epsfbox[95 109 648 506]{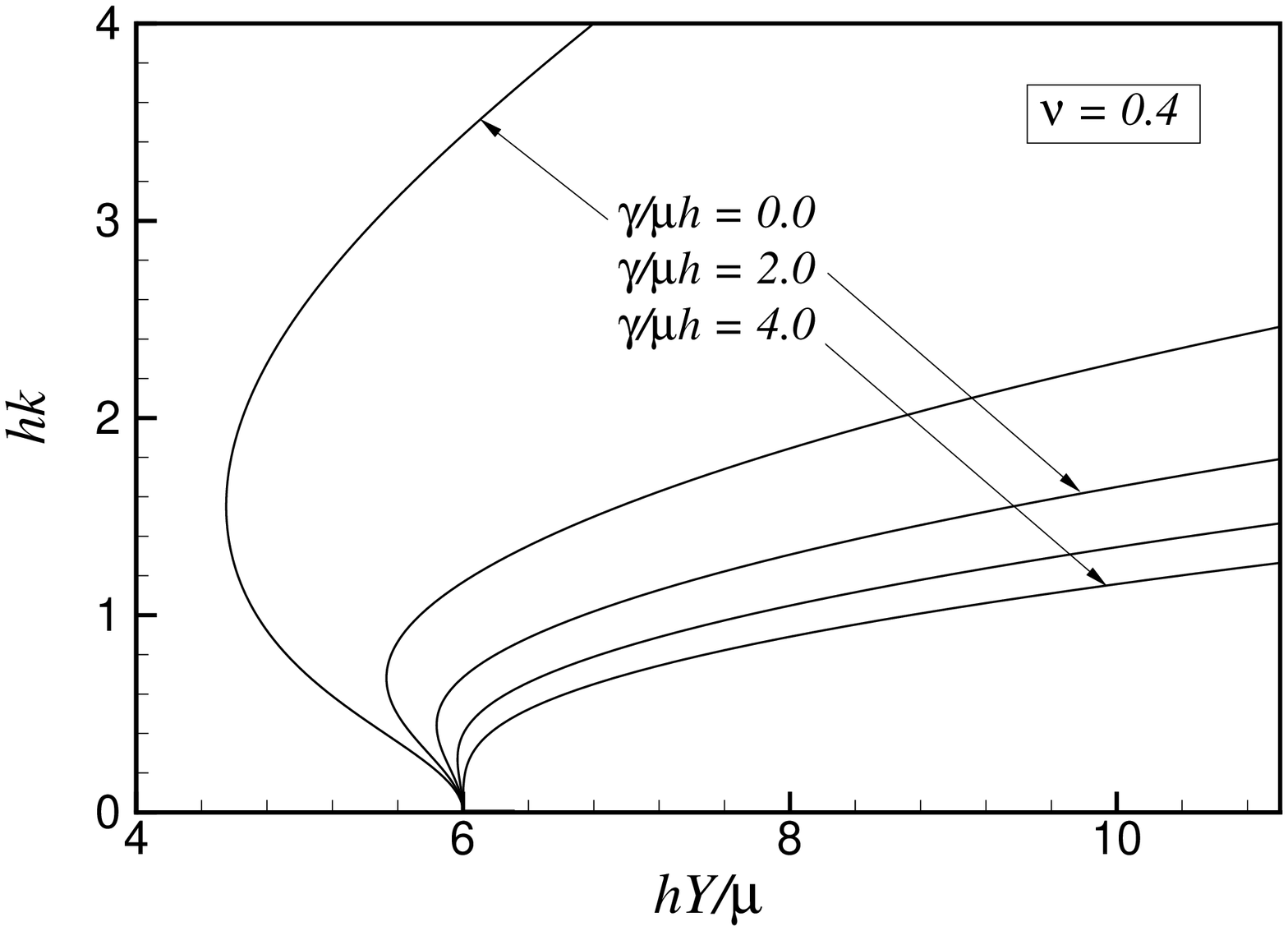}}
\centerline{(b)}
\caption{Bifurcations modes $(hk)$ as a function of $hY/\mu$ for
various values of $\nu$ with $\gamma/\mu h = 0$. (b)Bifurcations modes
$(hk)$ as a function of $hY/\mu$ for various values of $\gamma/\mu h$
with $\nu = 0.4$. }
\label{yroot}
\vspace{-0.25truecm} 
\end{figure}
The equilibrium stress field $\bsig$ in the film (which minimises the
potential energy \prn{enefunc} over an appropriate length of the film)
satisfies the equilibrium equation $\nabla \cdot \bsig = \bf{0}$
in $V$ and the boundary condition
{\small 
\bea
\bsig \cdot \bn = \gamma u_{2,11} \bn + \Fo \bn + Y (\bu \cdot \bn) \bn \label{bcsurf}
\eea
}
on $S$. Taking the film to be an isotropic linear elastic solid with shear
modulus $\mu$ and Poisson's ratio $\nu$, gives a standard expression
for the strain energy density \cite{Landau1989} with a resulting
expression for the stress tensor expressed in terms of the gradient of
displacement. Thus the problem can be cast into a boundary value
problem for the unknown displacement field with the boundary
condition of vanishing displacements at $x_2 = -h$ at the the film
substrate interface in addition to \prn{bcsurf}.

\noindent
{\em The Homogeneous Solution:} A solution to the above boundary value problem exists such that the
stresses in the film are equal everywhere. This homogeneous solution $(\bu^h)$
is  $u^h_1 = 0$ everywhere, and $u^h_2$ has a linear
variation with $x_2$ starting from 0 at $x_2 = -h$, i.e., 
{\small \bea
  u_2^h(x_1,0)= \uo  = \frac{F_o}{\left(\frac{2(1-\nu)
\mu}{(1-2\nu)h} - Y\right)}. \label{Uh}
\eea }
For the case when $\nu = 0.5$, i.~e., the incompressible limit, the
homogeneous solution is such that the displacement vanishes everywhere
in the film, and a pressure field $p$ develops such that $p(x_1,x_2)
= F_o$. So long as
{\small \bea
 Y < Y_m,\;\;\;\;\frac{h Y_m}{\mu} =  \frac{2(1-\nu)}{(1-2\nu)} \label{yhcond}
\eea }
the homogeneous solution is meaningful in that $\uo$ has the same sign
as $\Fo$. This conditions on $Y$ is most easily met when $\nu$ is
close to $0.5$ (the r.h.s.~of \prn{yhcond} tends to $\infty$ as $\nu$ tends to 0.5), i.~e., when the material in nearly incompressible. It is this
class of materials that the focus of this paper. Nevertheless,
results are presented for all values of $\nu$ for the sake of
completeness. 

\noindent
{\em Bifurcations:} What are the conditions (on $Y,\mu,\nu,h,\gamma$)
for another solution (inhomogeneous state) to exist? 
 If such a solution  exists, it can be taken to be of the
form $\bu^h + \bu$, where the symbol $\bu$ now stands for a
``bifurcation'' displacement field. This bifurcation field must
satisfy the equilibrium equations in the bulk and the rigid boundary
condition at the film substrate interface, just as the homogeneous
solution.  On the surface of the film at $x_2 = 0$, the bifurcation
field satisfies (here $\bsig$ is the {\em additional} stresses due to $\bu$),
{\small \bea
\bsig \cdot \bn & = &  \gamma u_{2,11} \bn + Y (\bu \cdot \bn) \bn,  \label{pbcsurf}
\eea }
instead of \prn{bcsurf}. 
To investigate the existence of a nontrivial solution to the problem
defined above, the bifurcation fields are assumed to have the form
{\small \bea
u_j(x_1,x_2) = e^{i k x_1}u_j(x_2) \label{pfields}
\eea }
where $k$ is a real positive wavenumber. The problem of finding
nontrivial bifurcation fields can be cast into the problem of finding
those values of $k$ such that the functions $u_j(x_2)$ are
nontrivial. It can be shown that (a detailed account will be published
elsewhere) nontrivial bifurcation fields of the form \prn{pfields} exist for those values of $k$
that satisfy the equation

{\small \bea
& & \Big( k \left[ 4 e^{2hk} h k^2 \left(h \mu - (1 - \nu) \gamma\right) +
(e^{4hk} -1) k \gamma ( 3 - 7 \nu + 4 \nu^2) \right.  \non \\
& & \left.  + \mu \left( (3 - 4 \mu) (1 + e^{4hk}) - 2 e^{2hk}(5 - 12 \nu + 8
\nu^2)\right)\right] \Big) \Big/ \non \\
& & \Big((1 - \nu)\left[ (3 - 4 \nu)(e^{4hk} - 1) - 4 hk e^{2 h k}
\right] \Big)  = Y \label{keqn}
\eea }
\noindent
This relation is valid for the incompressible case as well (i.~e., when $\nu
= 0.5$).  Real roots of \prn{keqn} are sought when $Y < Y_m$ which is
the range of $Y$ for which the homogeneous solution is valid.

We first focus attention on the case when $\gamma$
vanishes. \Fig{yroot}a depicts graphically the solution to \prn{keqn},
i.~e., for a given value of $\nu$, the values of $k$ that solve
\prn{keqn} are plotted as a function of $Y$ ($hY/\mu$ in
non-dimensional terms).  The important results may be noted: (i) There
are no bifurcation modes for any value of $\nu$ when $hY/\mu <
2$. (ii) For all values of $\nu$, $k=0$ is a bifurcation mode when $Y
= Y_m$. (iii)When $\nu \le 0.25$, there are no bifurcation modes for
$Y < Y_m$. (iv) When $\nu > 0.25$, there are two modes starting from a
critical value $Y_c$ (such as the point $C$ shown in \fig{yroot}a)
that depends on the value of $\nu$ until $Y$ reaches $Y_m$. When the
film is incompressible $h Y_c/\mu = 6.22$ and the corresponding
bifurcation mode has $h k_c = 2.12$. For this case bifurcations are
possible for all values of $Y$ greater than $6.22 \mu /h$, with two
possible values of $k$ as shown in the \fig{yroot}a.

Next, we consider the case when $\gamma \ne 0$. \Fig{yroot}b shows a
plot of the  possible wavenumbers of bifurcation modes for various values of $\gamma$ with $\nu = 0.4$. The key
effect of the surface energy  on the bifurcation modes are noted as
follows: (i) Surface energy inhibits bifurcation, in that a larger value of
$Y_c$ is effected with a non zero value of $\gamma$. The critical
mode $k_c$ decreases with increasing $\gamma$. Both of these results
are as expected since a larger value of $k$ implies a larger energy
penalty in terms of surface energy. (ii) As $\gamma$ gets
larger $Y_c$ approaches $Y_m$. In fact, it can be shown that $Y_c$
equals $Y_m$ when $\gamma = \gamma_m$ where
{\small \bea
\frac{\gamma_m}{\mu h} = \frac{2 \nu (4 \nu -1)}{3 (1-2\nu)^2}, \label{gmax}
\eea }
a result which is pertinent when $\nu > 0.25$. The curve for
$\gamma/\mu h = 4.0$ for the case of $\nu = 0.4$ shown in \fig{yroot},
graphically illustrates this point. If $\gamma > \gamma_m$, then there
are no bifurcations in the physically meaningful range $Y < Y_m$.

A more detailed analysis gives the following formulae for $Y_c$ and
$k_c$ as a function of $\gamma$ and $\nu$ when $\gamma/\mu h \ll 1$
and $\nu \rightarrow 0.5$:
{\small \bea
\frac{h}{\mu} Y_c(\nu,\gamma/\mu h) & = & 6.22 - 10.46 (1 - 2 \nu) + 4.49
\frac{\gamma}{\mu h}, \non \\ 
 h k_c(\nu,\gamma/\mu h) & = & 2.12 - 2.86 (1- 2 \nu) - 2.42
\frac{\gamma}{\mu h}. \label{yceqn}
\eea }

It is also interesting to consider the time evolution of deformation
in the film so as to obtain the dominant or the fastest growing mode. To this end, the film is considered to
be viscoelastic with a constitutive relation of the form
{\small \bea
\bsig & = & 2 \mu \left(\half (\nabla \bu + \nabla \bu^T) + \frac{\nu}{1 -
2 \nu} \nabla \cdot \bu \bI \right) \non \\ 
& & + 2 \eta \left(\half(\nabla
\dot{\bu} + \nabla \dot{\bu}^T  ) - \frac{1}{3}\nabla \cdot
\dot{\bu} \bI \right), \label{vconseqn}
\eea }
where $(\dot{})$ stands for the time derivative, $\eta$ is a
viscosity parameter and $\bI$ is the second order identity tensor. In
the consideration of the time evolution of the system, inertial
effects are neglected since the time scale of interest is much
larger than the time scale of the propagation of an elastic wave
through the thickness of the film.

\noindent
{\em The Homogeneous Viscoelastic Solution:} The homogeneous solution
of the field equations with the viscoelastic constitutive relation
\prn{vconseqn} is
{\small \bea
\!\!\!\!\!\!\!\! u^h_1 = 0, \;\; u^h_2(x_1,x_2,t) = \uo \left(1 +
\frac{x_2}{h}\right) \left( 1 - e^{\omega^h t} \right)
\eea }
where $\omega^h$ is given by
{\small \bea
\omega^h = -\frac{3}{4 \eta} \left(\frac{2(1-\nu)
\mu}{(1-2\nu)h} - Y \right) \label{omegah} = - \frac{3}{4
\eta}\left( Y_m - Y \right).
\eea }
From \prn{omegah} it is evident that the time dependent homogeneous
solution tends to the elastic homogeneous solution as $t \rightarrow
\infty$ when $Y < Y_m$. If $Y > Y_m$, the present analysis indicates
that the homogeneous solution blows up as  $t \rightarrow \infty$.

\begin{figure}
\centerline{\hspace{-1.0truecm}\epsfxsize=7.5truecm  \epsfbox[81 88 652 486]{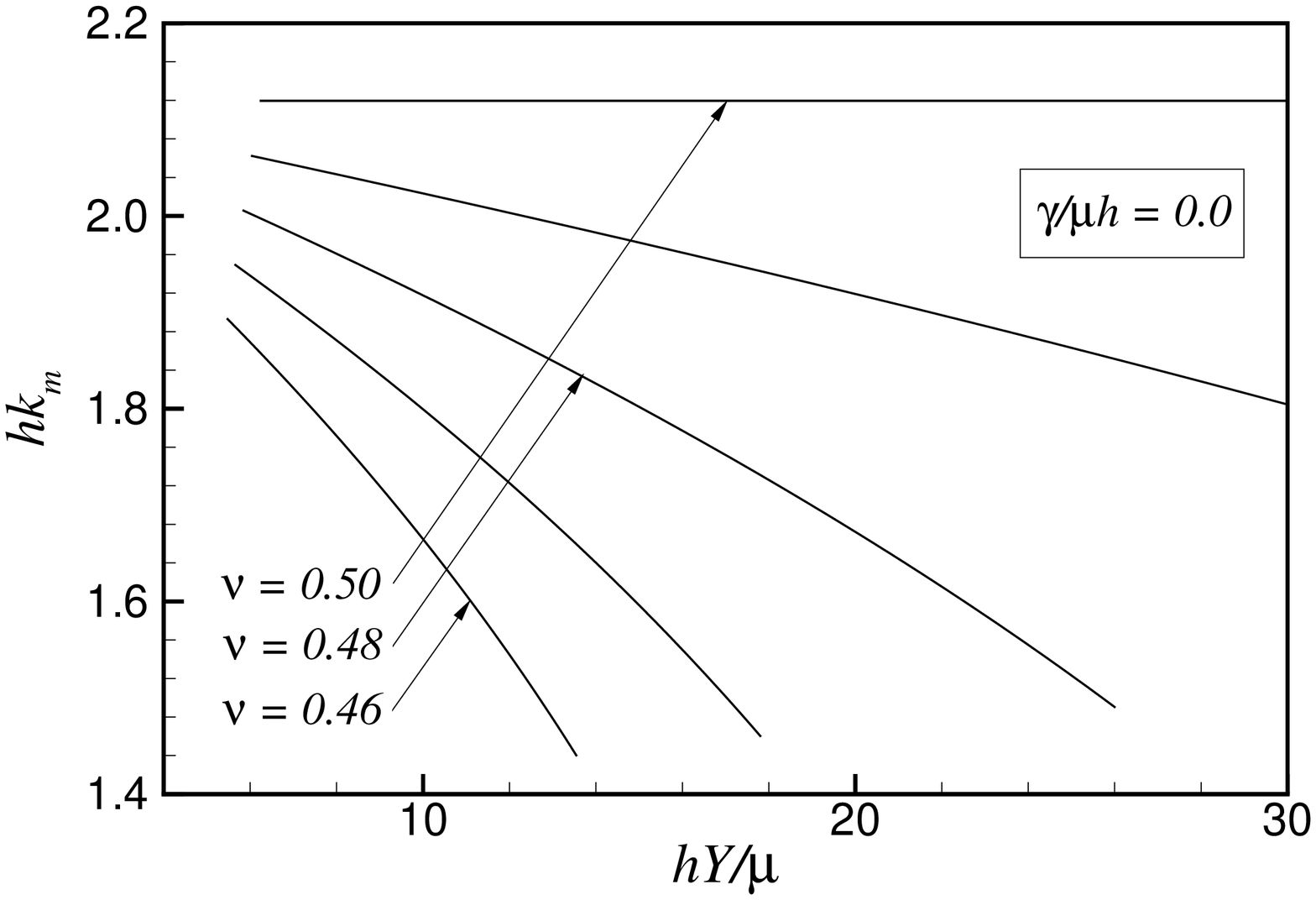}}
\centerline{(a)}
\centerline{\hspace{-1.0truecm}\epsfxsize=7.5truecm \epsfbox[81 88 652 486]{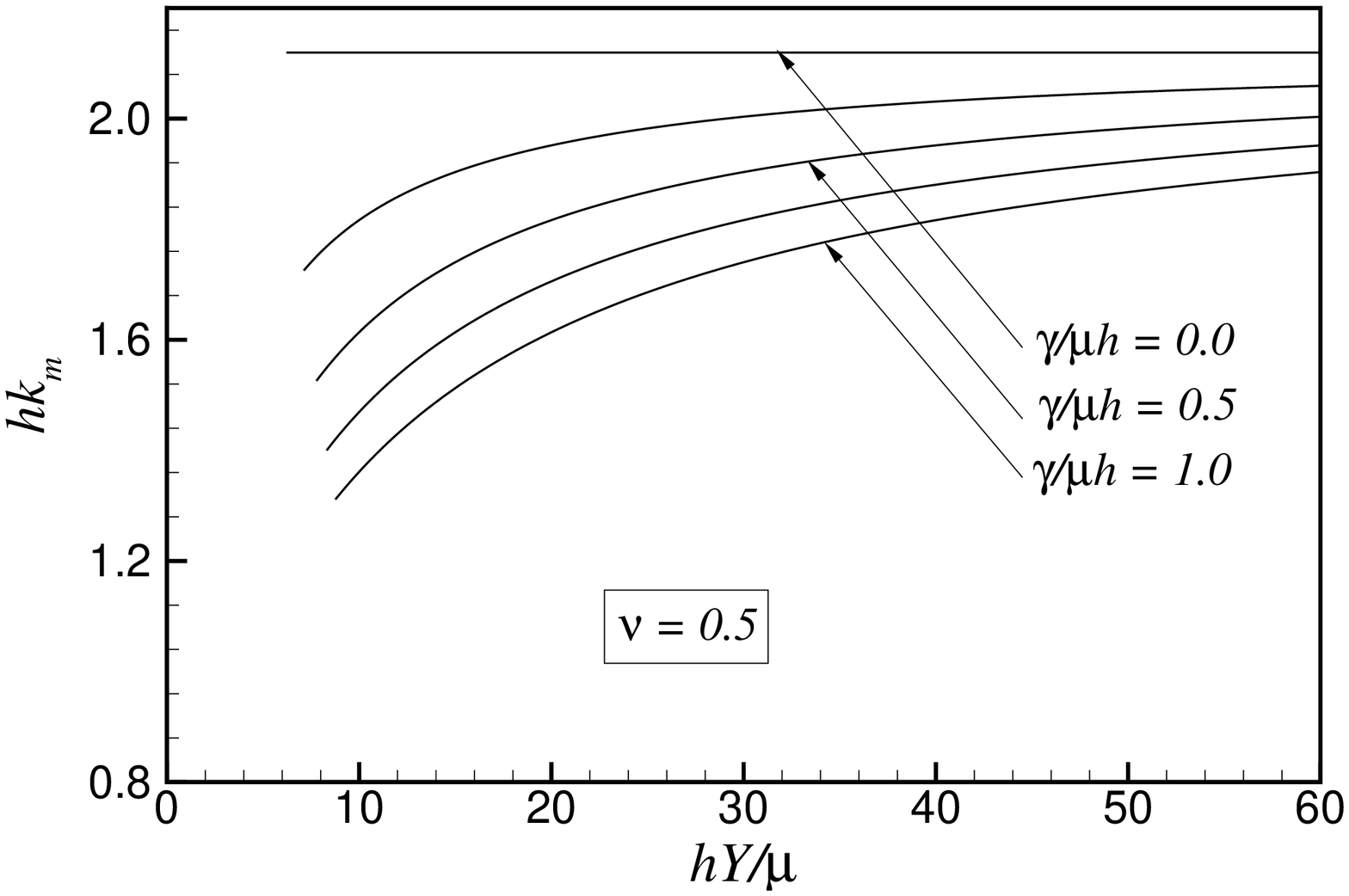}}
\centerline{(b)}
\caption{Bifurcations modes $(hk)$ as a function of $hY/\mu$ for
various values of $\nu$ with $\gamma/\mu h = 0$. (b)Bifurcations modes
$(hk)$ as a function of $hY/\mu$ for various values of $\gamma/\mu h$
with $\nu = 0.4$. }
\label{kmax}
\vspace{-0.25truecm} 
\end{figure}

\noindent
{\em Growth of Perturbations:} Just as in the case of the elastic
film, it is of interest to investigate the growth of perturbations of
the homogeneous solution. The perturbations $\bu$ are assumed to be of
the form
{\small \bea
u_j(x_1,x_2,t) = e^{ikx_1}u_j(x_2)e^{\omega t}.
\eea }
For a given $k$, the rate of growth $\omega$ is determined by
insisting that the the perturbation satisfies equilibrium equations and
boundary conditions and that they be nontrivial. The relation between
$\omega$ and $k$ can be obtained by replacing $\mu$ and $\nu$ in
\prn{keqn} respectively by $\mu^*$ and $\nu^*$ where,
{\small \bea
\mu^* = \mu + \eta \omega, \;\;\;\;\; \nu^* = \frac{3 \nu \mu - (1 -2
\nu)\eta \omega}{3 \mu + (1 - 2\nu) \eta \omega}.
\eea }
This procedure results in a cubic equation for $\omega$. The
solution of this equation is obtained by numerical means.

The solution for $\omega$ indicates that for $Y_c < Y < Y_m$, all
perturbation modes with wavenumbers between the two bifurcation modes
given by the elastic analysis are unstable i.~e., $\omega$ for these
modes are positive. Indeed, there is a mode
with wavenumber ($k_m$) between wavenumbers of the two elastic bifurcation
modes such that the rate of growth ($\omega$) is a
maximum. \Fig{kmax}(a) shows a plot of $k_m$ as a function of $Y$
($Y_c \le Y \le Y_m)$ for various values of $\nu$ (with $\gamma/\mu h
= 0)$. When $\nu < 0.5$, the value of $k_m$ starts at $k_c$ when $Y =
Y_c$ and monotonically falls with increasing $Y$. For the case of $\nu
= 0.5$, $k_m = k_c$ for all values of $Y$. When $\gamma \ne 0$, $k_m$
is smaller as is evident from \fig{kmax}(b); the effect of surface
energy on the fastest growing mode becomes increasingly less
significant for large values of $Y$. Just as in \prn{yceqn}, an
analytic result can be derived for $k_m$ for small values of
$\gamma/\mu h$, $\nu \rightarrow 0.5$ and $h(Y - Y_c)/\mu \ll 1$:
{\small \bea
h k_m(\nu, \frac{\gamma}{\mu h}) & = & hk_c(\nu, \frac{\gamma}{\mu h})
\non \\ & & \!\!\!\!\!\!\!\! \!\!\!\!\!\!\!\!\!\!\!\!\!\!\!\! + \left( 0.39 \frac{\gamma}{\mu h}-
0.46(1-2\nu)\right)\frac{h}{\mu} (Y - Y_c)\label{kmeqn}
\eea }

\begin{figure}
\centerline{\hspace{-1.0truecm}\epsfxsize=7.5truecm \epsfbox[93 106 658 502]{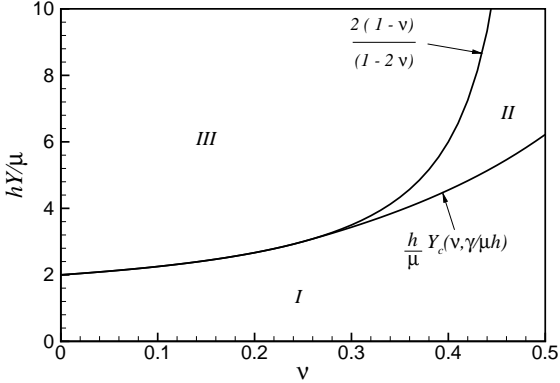}}
\caption{The stable and unstable regions in the parameter
space. Region $I$: Homogeneous solution stable. Region $II$:
Homogeneous solution unstable with two elastic bifurcation modes and a
fastest growing viscoelastic perturbation. Region $III$: Homogeneous
solution ``unphysical''.}
\label{regions}
\vspace{-0.25truecm}
\end{figure}

Instability in a thin film whose surface experiences forces depends on
three key sets of non-dimensional parameters namely the Poisson's
ratio $\nu$, the normalised second derivative of the interaction potential
$hY/\mu$ and the normalised surface energy $\gamma/\mu h$. The whole
picture of stability and bifurcation in this  system and its
dependence on the nondimensional parameters can be depicted
pictorially as shown in \fig{regions}.
Region $I$ in \fig{regions} is where the homogeneous solution is unique
and stable while region marked $III$ in the figure corresponds to the
case when the homogeneous solution is ``unphysical'', i.e., this
analysis is not adequate. Region $II$ is the most interesting -- this
corresponds to nearly incompressible material behaviour. In this
region the homogeneous solution is unstable, with two possible elastic
bifurcation modes; a viscoelastic analysis predicts a fastest growing
mode with a wave vector that lies between the two elastic bifurcation
modes.

We now turn to specific cases of the type of system considered in this
paper. First, we consider a rigid contactor interacting with the film via
van der Waals forces. Assuming that the contactor is at a distance $d$
above the undeformed surface of the film, the interaction potential
$U$ can can be taken to be $U(\bu \cdot \bn) = \frac{A}{12\pi (\bu
\cdot \bn - d)^2}$ with $\Fo =  \frac{A}{6 \pi d^3}$, $Y = \frac{A}{2
\pi d^4}$. Taking the film to be made of rubber ($\mu = 1$ MPa, $\nu =
0.5$, $\gamma = 0.1$J/m$^2$) and $h = 1$micron with  $A \approx 1$eV. When $d = 10$nanometers
we get $h Y/\mu = 1.6$ and for $d = 5$nanometers, $hY/\mu =
25.6$. Since the latter value is greater than $hY_c/\mu$ which is
6.63 when $\gamma/\mu h = 0.1$ (which is the present case), it is
clear that the condition for bifurcation will be achieved as $d$ is
reduced from $10$nm to $5$nm. Thus as the contactor approaches the
film, the film would buckle. This implies that the contact that forms
between the contactor surface and the film will not be planar. We are
not aware of any experimental work that can corroborate our
results. We do, however, hope that the contents of this paper will be
useful in designing experiments to verify our conclusions.

The second case considered is that of a film interacting with an
external electric field. The system consists of two plates separated
by a distance $d$; the bottom plate is coated with a nearly
incompressible polymeric film of height $h$. A potential difference
of $V$ is applied between the two plates. The quantity of interest is
the value of the gap thickness $d-h$ at which instability occurs in
the film. The potential of interaction for this case is given by
$U(\bu \cdot \bn) = \frac{\varepsilon_0 \varepsilon_p  V^2}{2(
\varepsilon_p d - (\varepsilon_p - 1) (h + \bu \cdot \bn))}$ where
$\varepsilon_0$ is the permittivity of free space, $\varepsilon_p$ is
the dielectric constant of the polymer. Taking the mechanical
properties of the polymer to be same as in the previous case, and
taking $\varepsilon_p = 3$, we get that the critical gap thickness
$d-h$ of $0.05$micron for a film of height $0.1$micron with the
applied voltage of 100V. A gap thickness smaller than 0.05micron will
cause the film to buckle. It is evident that large electric fields are
required to cause the instability.

We do wish to point out that this analysis is based on a linearised
model, and will only provide the modes of instability, i.~e., the
wavelength of surface undulation and not the magnitude. A nonlinear
analysis is required to obtain such a quantity and will be pursued in
subsequent papers.
\vspace{-0.5truecm}
\bibliography{sfb}
\end{document}